\documentclass[aps,prb,twocolumn,superscriptaddress,floatfix]{revtex4-1}

\usepackage{makecell}
\usepackage{graphicx}
\usepackage{dcolumn}
\usepackage{bm}
\usepackage{times}
\usepackage{color}
\usepackage{appendix}
\usepackage{ulem}

\begin{document}
	\bibliographystyle{apsrev4-1}
\title{Quantum oscillations and weak anisotropic resistivity in the chiral Fermion semimetal PdGa}
	
\author{Xiang-Yu Zeng}\thanks{These authors contributed equally to this paper}
\affiliation{Department of Physics, Renmin University of China, Beijing 100872, P. R. China}
\affiliation{Beijing Key Laboratory of Opto-electronic Functional Materials $\&$ Micro-nano Devices, Renmin University of China, Beijing 100872, P. R. China}
\author{Zheng-Yi Dai}\thanks{These authors contributed equally to this paper}
\affiliation{Department of Physics, Renmin University of China, Beijing 100872, P. R. China}
\affiliation{Beijing Key Laboratory of Opto-electronic Functional Materials $\&$ Micro-nano Devices, Renmin University of China, Beijing 100872, P. R. China}
\author{Sheng Xu}\thanks{These authors contributed equally to this paper}
\affiliation{Department of Physics, Renmin University of China, Beijing 100872, P. R. China}
\affiliation{Beijing Key Laboratory of Opto-electronic Functional Materials $\&$ Micro-nano Devices, Renmin University of China, Beijing 100872, P. R. China}
\author{Ning-Ning Zhao}\thanks{These authors contributed equally to this paper}
\affiliation{Department of Physics, Renmin University of China, Beijing 100872, P. R. China}
\affiliation{Beijing Key Laboratory of Opto-electronic Functional Materials $\&$ Micro-nano Devices, Renmin University of China, Beijing 100872, P. R. China}

\author{Huan Wang}
\affiliation{Department of Physics, Renmin University of China, Beijing 100872, P. R. China}
\affiliation{Beijing Key Laboratory of Opto-electronic Functional Materials $\&$ Micro-nano Devices, Renmin University of China, Beijing 100872, P. R. China}
\author{Xiao-Yan Wang}
\affiliation{Department of Physics, Renmin University of China, Beijing 100872, P. R. China}
\affiliation{Beijing Key Laboratory of Opto-electronic Functional Materials $\&$ Micro-nano Devices, Renmin University of China, Beijing 100872, P. R. China}
\author{Jun-Fa Lin}
\affiliation{Department of Physics, Renmin University of China, Beijing 100872, P. R. China}
\affiliation{Beijing Key Laboratory of Opto-electronic Functional Materials $\&$ Micro-nano Devices, Renmin University of China, Beijing 100872, P. R. China}
\author{Jing Gong}
\affiliation{Department of Physics, Renmin University of China, Beijing 100872, P. R. China}
\affiliation{Beijing Key Laboratory of Opto-electronic Functional Materials $\&$ Micro-nano Devices, Renmin University of China, Beijing 100872, P. R. China}
\author{Xiao-Ping Ma}
\affiliation{Department of Physics, Renmin University of China, Beijing 100872, P. R. China}
\affiliation{Beijing Key Laboratory of Opto-electronic Functional Materials $\&$ Micro-nano Devices, Renmin University of China, Beijing 100872, P. R. China}
\author{Kun Han}
\affiliation{Department of Physics, Renmin University of China, Beijing 100872, P. R. China}
\affiliation{Beijing Key Laboratory of Opto-electronic Functional Materials $\&$ Micro-nano Devices, Renmin University of China, Beijing 100872, P. R. China}
\author{Yi-Ting Wang}
\affiliation{Department of Physics, Renmin University of China, Beijing 100872, P. R. China}
\affiliation{Beijing Key Laboratory of Opto-electronic Functional Materials $\&$ Micro-nano Devices, Renmin University of China, Beijing 100872, P. R. China}
\author{Peng Cheng}
\affiliation{Department of Physics, Renmin University of China, Beijing 100872, P. R. China}
\affiliation{Beijing Key Laboratory of Opto-electronic Functional Materials $\&$ Micro-nano Devices, Renmin University of China, Beijing 100872, P. R. China}

\author{Kai Liu}\email{tlxia@ruc.edu.cn}\email{kliu@ruc.edu.cn}
\affiliation{Department of Physics, Renmin University of China, Beijing 100872, P. R. China}
\affiliation{Beijing Key Laboratory of Opto-electronic Functional Materials $\&$ Micro-nano Devices, Renmin University of China, Beijing 100872, P. R. China}
\author{Tian-Long Xia}\email{tlxia@ruc.edu.cn}\email{kliu@ruc.edu.cn}
\affiliation{Department of Physics, Renmin University of China, Beijing 100872, P. R. China}
\affiliation{Beijing Key Laboratory of Opto-electronic Functional Materials $\&$ Micro-nano Devices, Renmin University of China, Beijing 100872, P. R. China}

\date{\today}

\begin{abstract}
We perform a detailed analysis of the magnetotransport and de Haas-van Alphen (dHvA) oscillations in crystal PdGa which is predicted to be a typical chiral Fermion semimetal from CoSi family holding a large Chern number. The unsaturated quadratic magnetoresistance (MR) and nonlinear Hall resistivity indicate that PdGa is a multi-band system without electron-hole compensation. Angle-dependent resistivity in PdGa shows weak anisotropy with twofold or threefold symmetry when the magnetic field rotates within the (1$\bar{1}$0) or (111) plane perpendicular to the current. Nine or three frequencies are extracted after the fast Fourier-transform analysis (FFT) of the dHvA oscillations with B//[001] or B//[011], respectively, which is confirmed to be consistent with the Fermi surfaces (FSs) obtained from first-principles calculations with spin-orbit coupling (SOC) considered.
\end{abstract}
\maketitle

The discovery of topological materials has kick-started a revolution in the field of condensed matter physics. Dirac and Weyl semimetals have been widely studied due to their novel properties\cite{wehling2014dirac,neupane2014observation,liu2014stable,borisenko2014experimental,liang2015ultrahigh,li2015giant,li2016negative,wang2012dirac,liu2014discovery,xiong2015evidence,xiong2016anomalous,wan2011topological,fang2003anomalous,weng2015weyl,xu2015discovery,huang2015weyl,lv2015observation,xu2015experimental,xu2015discovery2,liu2016evolution,huang2015observation,zhang2016signatures,arnold2016negative}, such as high mobility, low carrier concentration, large magnetoresistance (MR), chiral anomaly induced negative magnetoresistance (NMR), \textit{etc.} In addition to the well-known inversion-symmetry-broken Weyl semimetal TaAs and its family members which hold spin-1/2 Weyl fermion\cite{lv2015experimental,lv2015observation,xu2015experimental,xu2016observation,liu2016evolution,huang2015observation,zhang2016signatures,arnold2016negative}, some new types of fermionic excitations have been proposed to possess large Chern numbers, namely spin-1 excitation\cite{manes2012existence,tang2017multiple,bradlyn2016beyond}, double Weyl fermion\cite{tang2017multiple,xu2016type} and spin-3/2 Rartia-Schwinger-Weyl (RSW) fermion\cite{liang2016semimetal,ezawa2016pseudospin,rarita1941theory} \textit{etc.} Recently, lots of significant work on CoSi, RhSi and RhSn has confirmed the existence of these predicated fermions\cite{pshenay2018band,sanchez2019topological,rao2019observation,takane2019observation,xu2019crystal,wu2019single,chang2017unconventional,xu2019quantum,wang2020haas}, which brings a further deep understanding of the chiral fermions. These materials hold the spin-1 excitation at $\Gamma$, double Weyl fermion at R in the first Brillouin zone with the Chern number $\pm$2, and two Fermi arcs connect the Weyl points on the surface states (SSs) without spin-orbit coupling (SOC). The band splitting might be obvious once the SOC is included. Furthermore, the spin-1 excitation and double Weyl fermion evolve into spin-3/2 RSW fermion and time-reversal (TR) doubling of the spin-1 excitation with the Chern number $\pm$4, respectively. Meanwhile, the two Fermi arcs split, which is hardly observed in ARPES due to the weak SOC in CoSi and RhSi. Recently, ARPES results on PdGa have comfirmed the existence of the predicted topological chiral surface Fermi arcs and band splitting due to the relatively stronger SOC\cite{schroter2020observation}. Meanwhile, detailed study on the magnetotransport properties in PdGa is lacking.

In this work, we grew the high quality single crystals of PdGa, measured the transport properties and studied its electronic structure through de Haas-van Alphen (dHvA) effect. Similar band structure and properties with its family materials CoSi or PtGa have been obtained while detailed difference on the Fermi surface structure exist. PdGa holds spin-1 excitation at $\Gamma$, double Weyl fermion at R in the first Brillouin zone without SOC, carrying the Chern number $\pm$2 . The band splits when SOC is considered, which makes the spin-1 excitation and  double Weyl fermion evolve into spin-3/2 RSW fermion and TR doubling of the spin-1 excitation with the Chern number $\pm$4, respectively. The transverse magnetoresistance of PdGa shows unsaturated behavior and the Hall conductivity can be described appropriately with the two-band model, which indicates that PdGa is a multi-band system. The angle-dependent resistivity, an effective probe method for detecting highly symmetric bulk FSs, is employed to explore PdGa. The distinct dHvA oscillations have been observed and nine fundamental frequencies have been extracted after the fast Fourier-transform (FFT) analysis when the magnetic field is applied along [001] direction, which is  consistent with the first-principles calculations when SOC is considered. The light cyclotron effective masses are extracted from the fitting of thermal factor in the Lifshitz–Kosevich (LK) formula, indicating the possible existence of massless quasiparticles.
\begin{figure}[htbp]
	\centering
	\includegraphics[width=0.48\textwidth]{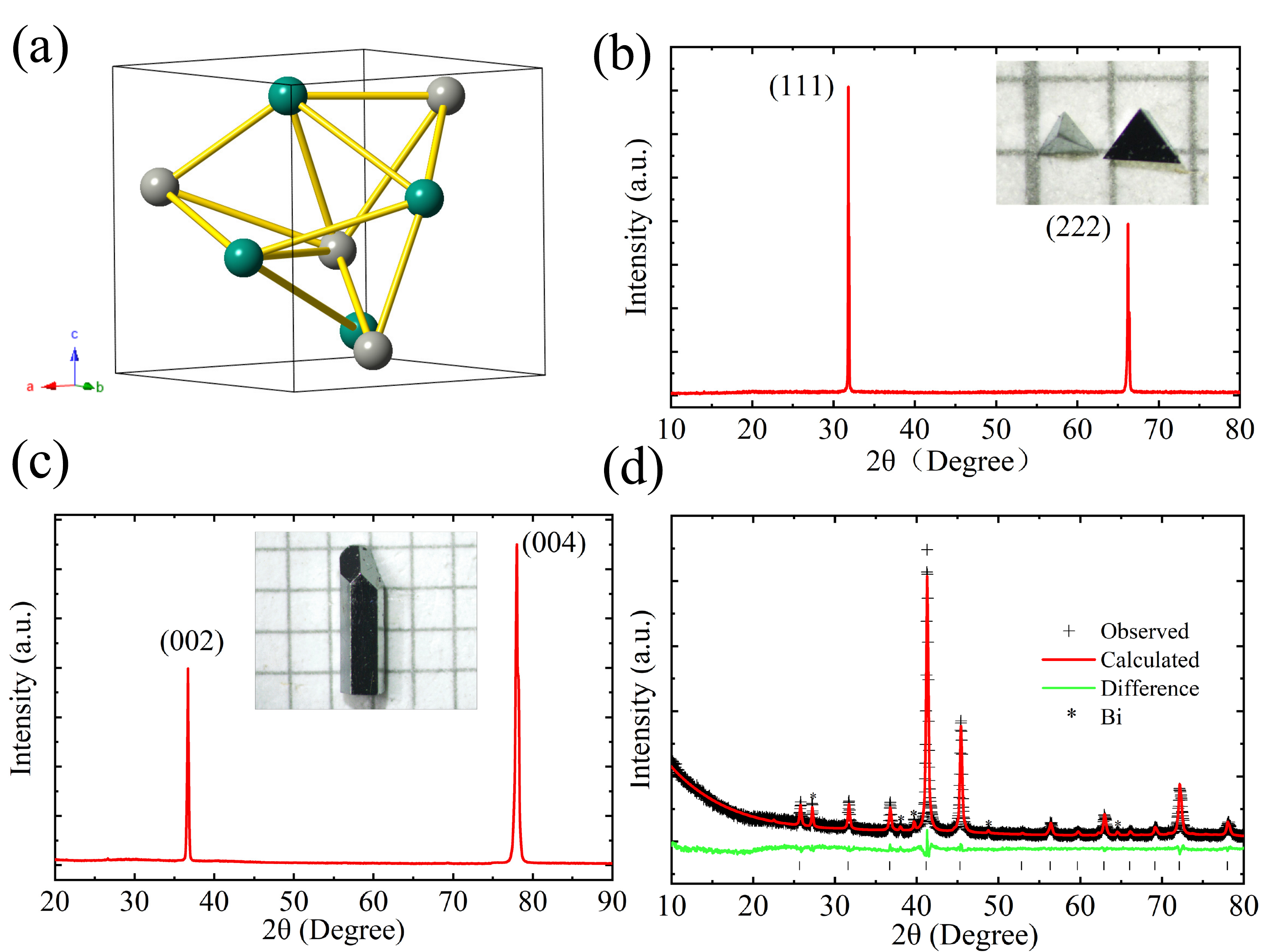}
	\caption{(a) Crystal structure of PdGa.  (b), (c) Single crystal XRD pattern of PdGa with (111) and (001). Insets show the picture of the grown crystal. (d) Powder XRD patterns and the Rietveld refinement of PdGa.}
\end{figure}

\section{Experimental and crystal structure}
The single crystals of PdGa were grown from bismuth flux. The starting elements, palladium powder, gallium ingot and excess bismuth granules were put into the corundum crucible and sealed in a quartz tube with a ratio of Pd: Ga: Bi=1:1:20. The quartz tube was heated to 950$^\circ$C at 60$^\circ$C/h and held for 20 h, then cooled to 400$^\circ$C at a rate of 1$^\circ$C/h, at which the excess Bi flux was separated from the crystals by centrifugation. The obtained crystals are triangular or rectangular shape with its size as large as 1$\sim$3 mm in one dimension. The atomic proportion of PdGa was checked to be Pd:Ga=1:1 using energy dispersive x-ray spectroscopy (EDS, Oxford X-Max 50). The single crystal and powder x-ray diffraction (XRD) were collected from a Bruker D8 Advance x-ray diffractometer using Cu K$_{\alpha}$ radiation. TOPAS-4.2 was employed for the refinement. The resistivity and Hall measurements were performed on a Quantum Design physical property measurement system (QD PPMS-14T) using the standard six-probe method on a long flake crystal. The magnetization was measured with the vibrating sample magnetometer (VSM) option of PPMS.

The first-principles electronic structure calculations on PdGa were performed by using the projector augmented wave (PAW) method\cite{blochl1994projector,kresse1999ultrasoft} as implemented in the VASP package\cite{kresse1993ab,kresse1996efficiency,kresse1996efficient}. The generalized gradient approximation (GGA) of Perdew-Burke-Ernzerhof (PBE) type\cite{perdew1996generalized} was used for the exchange-correlation functional. The kinetic energy cutoff of the plane-wave basis was set to 370 eV. The Brillouin zone was sampled with a $11\times11\times11$ k-point mesh. For the Fermi surface broadening, the Gaussian smearing method with a width of 0.05 eV was adopted. The lattice constants and the atomic positions were fully relaxed until the forces on all atoms were smaller than 0.01 eV/\AA. The relaxed lattice constants a=b=c=4.953(0)\AA \ are consistent with the experimental results. The spin-orbit coupling effect was included in the band structure calculations. The Fermi surfaces plane were calculated based on the tight-binding Hamiltonian constructed with the maximally localized Wannier functions\cite{mostofi2014updated,marzari2012maximally} for the outmost s, p and d orbitals of Pd atom, s, p and d orbitals of Ga atom generated by the first-principles calculations.

The crystal structure of PdGa is shown in Fig. 1(a), which crystallizes in a simple cubic structure with $P2_{1}3$ (No. 198) space group. Figures 1(b) and 1(c) display the single crystal XRD patterns with (111) and  (001) reflections. The insets of Figs. 1(b) and 1(c) exhibit pictures of the typical grown PdGa crystals with (111) and (001) crystal face, respectively. The powder XRD pattern is shown in Fig. 1(d), which can be indexed to the structure of PdGa with the refined lattice parameter a=b=c=4.89(0)\AA.

\section{Results and Discussions}

\begin{figure}[tbp]
	\centering
	\includegraphics[width=0.48\textwidth]{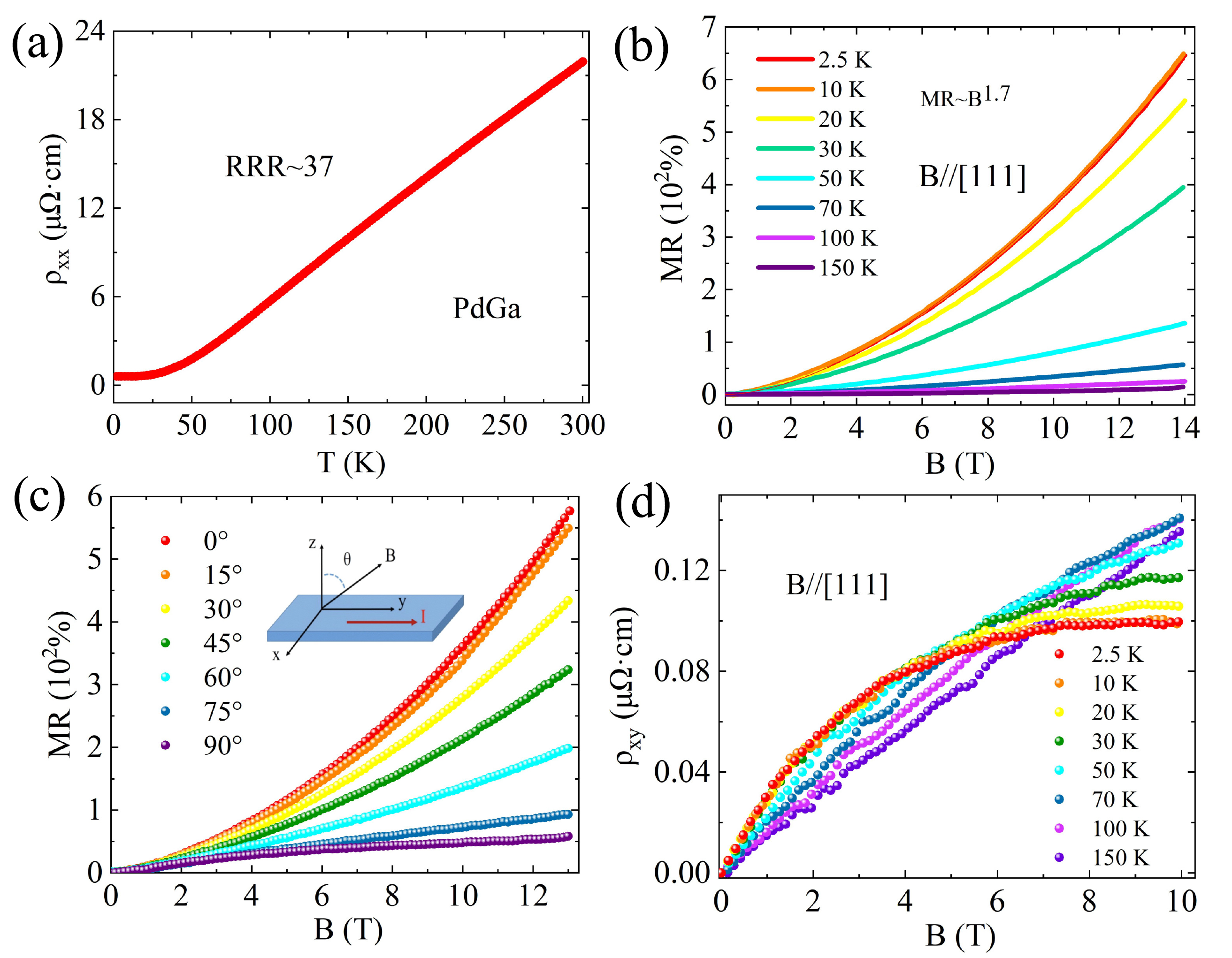}
	\caption{(a)Temperature dependence of the resistivity $\rho_{xx}$. (b) Magnetic field-dependent of MR at different temperatures. (c) Magnetic field-dependent MR at 2.5 K with magnetic field titled from B$\perp$I ($\theta$=$0^{\circ}$) to B//I ($\theta$=$90^{\circ}$). Inset shows the definition of $\theta$. (d)Magnetic field-dependent Hall resistivity at various temperatures.}
\end{figure}
\begin{figure}[b]
	\centering
	\includegraphics[width=0.5\textwidth]{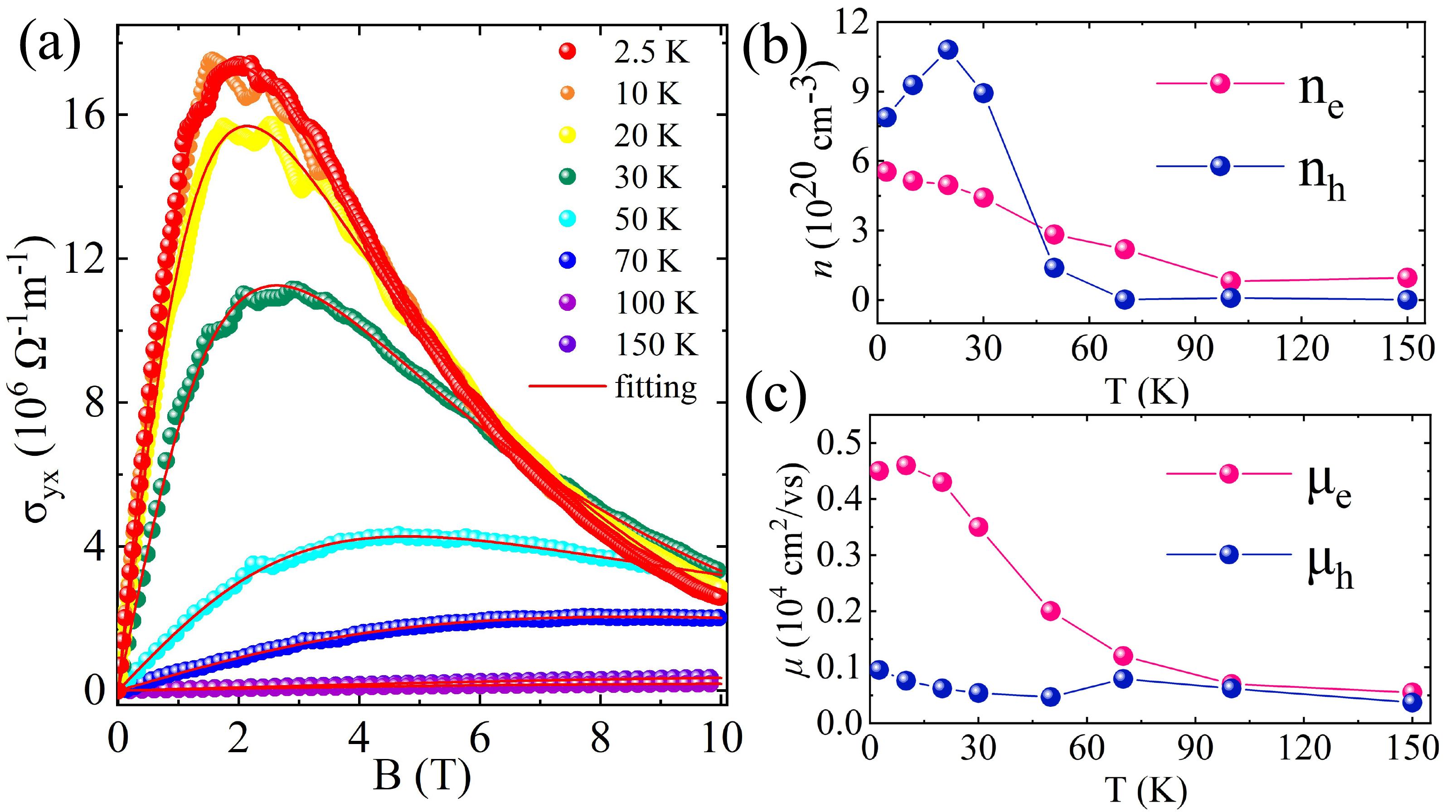}
	\caption{(a) Field dependence of the Hall conductivity $(\sigma_{y x}=\left.\rho_{x y} /\left(\left(\rho_{x y}\right)^{2}+\left(\rho_{x x}\right)^{2}\right)\right)$at various temperatures. (b) and (c) are the temperature dependence of carrier densities and mobility of the electrons and holes extracted from the two-band model fitting.}
\end{figure}
\begin{figure*}[t]
	\centering
	\includegraphics[width=\textwidth]{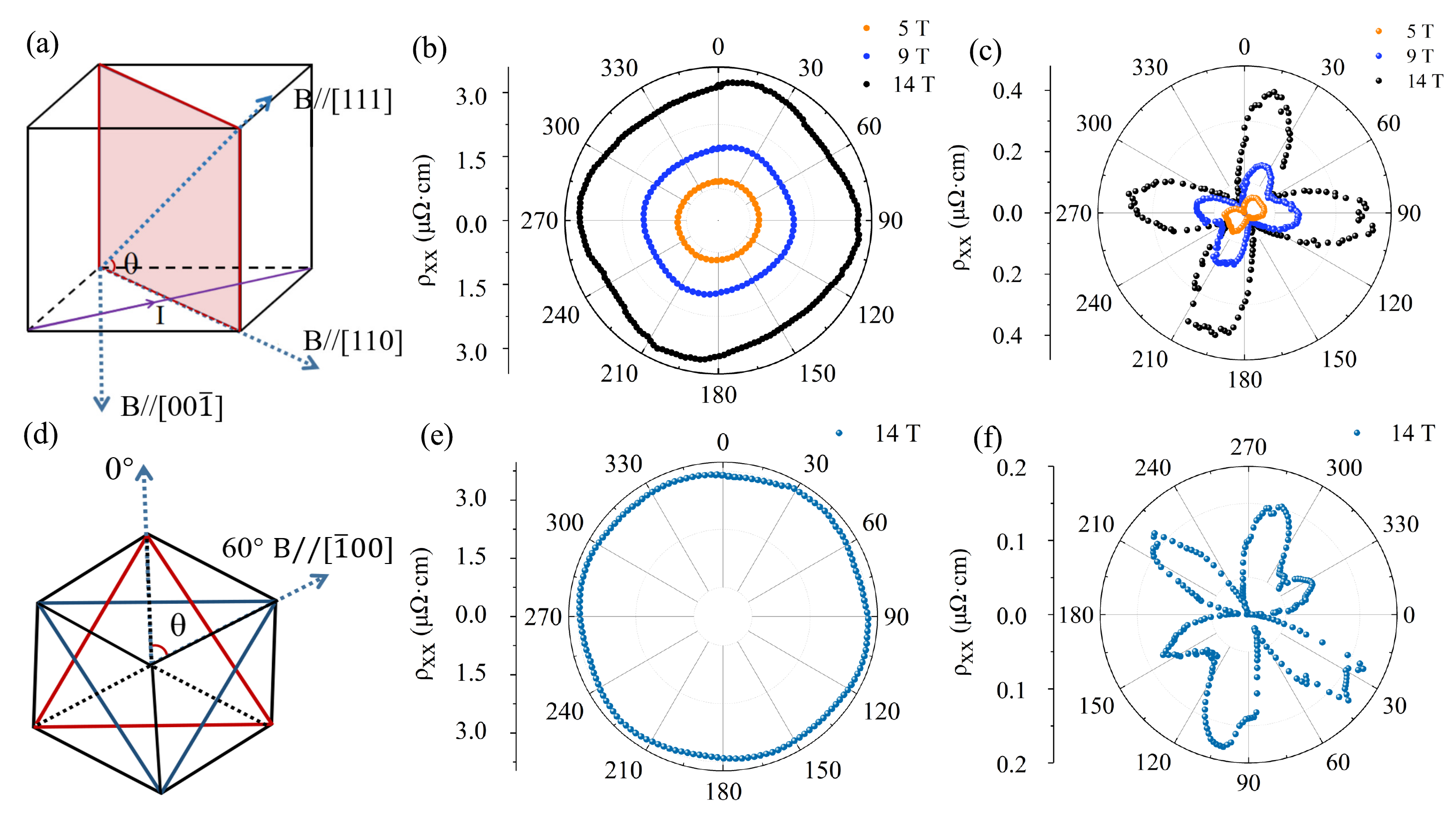}
	\caption{(a) and (d) show the schematic diagrams of angular rotation. (b) Polar plot of $\rho_{x x}$ under 5 T (yellow dots), 9 T (blue dots) and 14 T (black dots) with B vertical to I at 2.5 K. (c) Polar plot obtained from (b) after subtracting a minimum. (e) Polar plot of $\rho_{x x}$ under 14 T with B vertical to I at 2.5 K. (f) Polar plot obtained from (e) after subtracting a minimum.}
\end{figure*}
The temperature-dependent resistivity of PdGa is displayed in Fig. 2(a), which exhibits a metallic behavior with a relatively large residual resistivity ratio (RRR$\approx$37) indicating the high quality of the grown crystals. Fig. 2(b) presents the MR=$(\rho_{x y}(B)-\rho_{x y}(0))/\rho_{x y}(0)$ curves with B//[111] at various temperatures, which reaches about 600$\%$ at 2.5 K \& 14 T. As is shown, no sign of SdH oscillations was observed even the magnetic field is as large as 14 T. The field-dependent MR at 2.5 K with the magnetic field tilted from B$\perp$I ($\theta=0^{\circ}$) to B//I ($\theta= 90^\circ$) is exhibited in Fig. 2(c). The inset shows the definition of $\theta$ which is the angle between the magnetic field and current direction. The amplitude of MR decreases gradually with $\theta$ and reaches a minimum when B//I, where the NMR has not been observed due to the orbital MR induced by the trivial hole-like pockets at M or around $\Gamma$ point. In order to demonstrate the characteristics of  carriers in PdGa, we examine the evolution of the Hall resistivity $\rho_{x y}$ at various temperatures from 2.5 K to 150 K, as shown in Fig. 2(d). The Hall resistivity reveals a nonlinear behavior especially when the temperature decreases, indicating that PdGa is a multi-band system. Therefore, it is reasonable to fit the Hall conductivity with two-band model.

\begin{equation}\label{equ2}
	\centering
	\Delta \sigma_{y x}=\left(\frac{n_{e} \mu_{e}^{2}}{1+\left(\mu_{e} B\right)^{2}}-\frac{n_{h} \mu_{h}^{2}}{1+\left(\mu_{h} B\right)^{2}}\right) e B
\end{equation}

The $n_{e,h}$ and $\mu_{e,h}$ represent the concentration and mobility of electrons or holes, respectively. Fig. 3(a) displays the field-depentent Hall conductivity. The different color dots represent the experimental data while the red lines are fitting curves from which the temperature-dependent concentrations and mobilities are extracted, as shown in Figs. 3(b) and 3(c). As the temperature decreases, both the concentration and mobility of the two types of carriers increase. At 2.5 K, $n_{h}=7.9\times10^{20}cm^{-3}$, $n_{e}=5.5\times10^{20}cm^{-3}$ and $\mu_{h}$ = 950 $cm^{2}\cdot V^{-1}s^{-1}$, $\mu_{e}$ = 4500 $cm^{2}\cdot V^{-1}s^{-1}$, respectively.

\begin{table*}[!t]
	\centering
	\caption{Parameters derived from dHvA oscillations with B//[001]. F, oscillation frequency; A$_F$, cross sectional area of FS normal to the field; $m^*$, effective mass.}
	\label{oscillations}
	\setlength{\tabcolsep}{4.7mm}{
		\begin{tabular}{cccccccccc}
			\hline \hline & $F_{1}$ & $F_{2}$ & $F_{3}$ & $F_{4}$ & $F_{5}$ & $F_{6}$ & $F_{7}$ & $F_{8}$ & $F_{9}$ \\
			\hline$F(T)$ & 164.0 & 200.4 & 288.1 & 310.7 & 378.8 & 421.7 & 483.8 & 590.4 & 693.9 \\
			$m^{*} / m_{e}$ & 0.081 & 0.085 & 0.089 & 0.088 & 0.079 & 0.084 & 0.076 & 0.077 & 0.092  \\
			$A_{F} 10^{-3} \text{\AA}^{-2}$ & 15.63 & 19.12 & 27.50 & 29.66 & 36.15 & 40.26 & 46.18 & 56.35 & 66.23 \\
			\hline \hline
	\end{tabular}}
\end{table*}
\begin{figure*}[t]
	\centering
	\includegraphics[width=\textwidth]{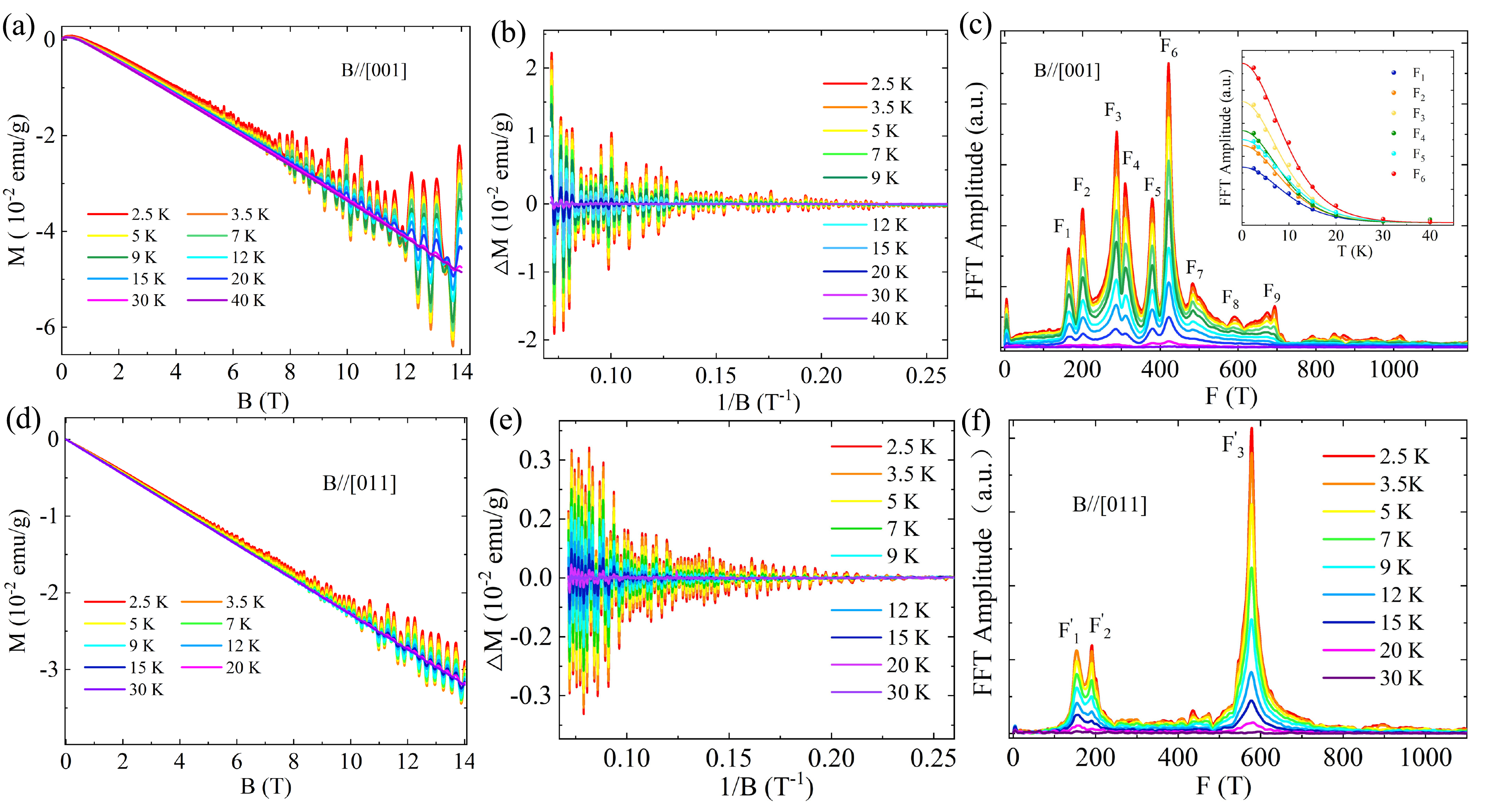}
	\caption{(a) and (d) are the dHvA oscillations at various temperatures with the B//[001] and B//[011] configuration respectively. (b) and (e) are the amplitudes of dHvA oscillations as a function of 1/B. (c) and (f) are the FFT spectras of the oscillations extracted from (b) and (e) respectively. Inset of (e) is the temperature dependence of relative normalized FFT amplitude of the frequencies.}
\end{figure*}

\begin{table}[tb]
	\centering
	\caption{Parameters derived from dHvA oscillations with B//[011]. F, oscillation frequency; $A_F$, cross sectional area of FS normal to the field.}
	\label{oscillations}
	\setlength{\tabcolsep}{2.1mm}{
		\begin{tabular}{cccc}
			\hline \hline & $F_{1}^{'}$ & $F_{2}^{'}$ & $F_{3}^{'}$ \\
			\hline $F(T)$ & 154.1 & 190.7 & 578.4 \\
			$A_{F} 10^{-3} \text{\AA}^{-2}$ & 14.70 & 18.19 & 55.19 \\
			\hline \hline
	\end{tabular}}
\end{table}

The anisotropy in the resistivity at different fields is presented in Fig. 4. A naturally grown regular tetrahedron sample with (111) plane is polished into a rectangular flake. Figure 4(a) shows the schematic diagram of the field rotation, where the current is applied along the [$\overline{1}$10] direction and the magnetic field is rotated within the ($\overline{1}$10) plane. At 2.5 K, the $\rho_{x x}$ at various fields are measured, as presented in Fig. 4(b), which shows a weak twofold symmetry that becomes more evident as the field increases. It should be noted that there exist a few data points overlapping at the beginning and the end of the rotation which are not exactly the same especially when a higher field is applied. As shown in Fig. 4(c), the field-dependent resistivity is normalized by subtracting the minimal value of the oscillating resistivity at the corresponding field in the polar plot to enhance the feature of the slight anisotropy. An analogous experiment is conducted on another sample with B rotating within (111) plane, as shown in Fig. 4(d). A cuboid with a cross section of 0.2 $\times$ 0.2 mm$^{2}$ was obtained after a reguler tetrahedron sample was polished and the current was applied along the direction of [111] (length of the sample) with the field rotating perpendicular to the current. Fig. 4(e) shows a polar plot of $\rho_{x x}$ at 14 T, which illustrates almost a circular shape. After subtracting a minimum from the raw data, six peaks with diverse intensities appear as exhibited in Fig. 4(f). These peaks can be divided into two groups with threefold symmetry, respectively, corresponding to the schematic diagram of Fig. 4(d). Regarding the symmetry, a classical theory describes\cite{zhu2012field,collaudin2015angle} that $\rho$ (B) is dominated by the orbital effects deriving from the cyclotron motion of the electrons caused by Lorentz force, corresponding to the formula\cite{xu2017angle} 
		\begin{equation}\label{equ2}
			\centering
			MR=\frac{\rho (B)-\rho (0)}{\rho (0)} \sim\left(w_{c} \tau\right)^{2} \sim\left(e \tau / m^{*}\right)^{2} B^{2} \sim \mu^{2} B^{2}
		\end{equation}
		\begin{figure*}[t]
			\centering
			\includegraphics[width=\textwidth]{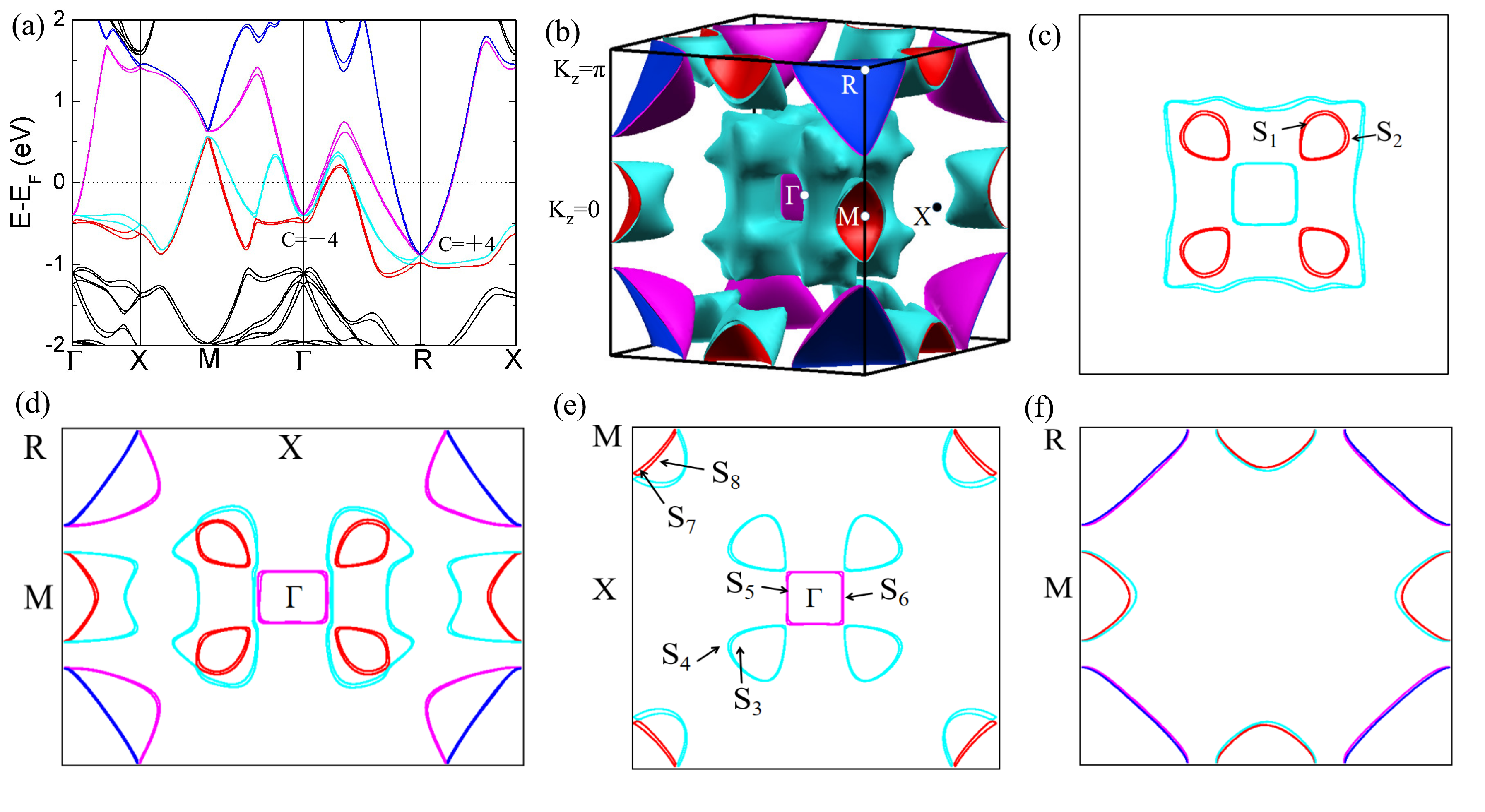}
			\caption{(a) Calculated bulk band structure of PdGa along highsymmetry lines with SOC. (b) Calculated FSs in the bulk Brillouin zone. (c) (d) (e) (f) Calculated FSs in k$_z$=0.16$\pi$ plane, $\Gamma$-X-M-R plane, k$_z$=0 plane and k$_z$=$\pi$ plane, respectively.}
		\end{figure*}
		where $w_{c}$ is the cyclotron frequency, $\tau$ is the scattering time, $m^*$ is the effective mass, and B is the magnetic field. Although the relaxation time varies along different directions, in the simplest model, the difference is so small that we can assume that $\tau$ is a constant. Therefore, the anisotropy of the effective masses of carriers strongly affect the values of the angle-dependent resistivity, which is directly related to the shape of the FSs. Different from the previous study on bismuth\cite{zhu2012field,collaudin2015angle}, where valley polarization is induced at high field and low temperature, resulting in symmetry breaking and anisotropic resistivity. The anisotropy in PdGa is relatively weak, the subtle difference of $\tau$ or a small misalignment between the magnetic field rotation plane and the current direction may both cause significant asymmetry in the resistivity so that the symmetry is broken. The nonnegligible anisotropy of PdGa reveals the three dimensional feature of Fermi surface and the differences of the carrier mobility and effective masses for fields along different directions.

Previous ARPES results present the band splitting of surface states with different chirality, signifying the existence of SOC in PdGa. On the other hand, quantum oscillation, usually an effective method to map the characteristics of FS according to the Onsager relation $F=\left(\phi_{0} / 2 \pi^{2}\right) A_{\mathrm{F}}=(\hbar / 2 \pi e) A_{\mathrm{F}}$, is employed in this work to study the electronic structure and SOC effect. The frequency F is proportional to the extreme cross section ($A_{F}$) of the FS normal to the magnetic field. As the field increases, the quantized Landau level will cross the Fermi energy ($E_{F}$) successivelly, which leads to the oscillation of the density of state (DOS) at $E_{F}$, eventually resulting in quantum oscillation on transport properties. Figures 5(a) and (d) present the isothermal magnetization of a prismatic crystal as shown in the inset of Fig. 1(c) with  B//[001] and B//[011] configurations, respectively, which both exhibit evident dHvA oscillations. The oscillatory components of magnetization are obtained after subtracting a smooth background as exhibited in Figs. 5(b) and (e).  After the FFT analysis, nine/three fundamental frequencies are extracted as shown in Figs. 5(c) and (f). Thus, according to the Onsager relation, the corresponding extreme cross section area of F$_{1}$-F$_{9}$ and F$^{'}_{1}$-F$^{'}_{3}$ with B//[001] and B//[011] are determined (see Table I and Table II for details). The low frequency at about 3 T in the FFT spectra comes from the data processing instead of the intrinsic dHvA oscillations. The oscillatory component versus 1/B is  described by the LK formula:
\begin{equation}\label{equ2}
	\centering
	\Delta M \propto -B^{1/2}\frac{\lambda T}{sinh(\lambda T)}e^{-\lambda T_D}sin[2\pi(\frac{F}{B}-\frac{1}{2}+\beta+\delta)]
\end{equation}
where $\lambda= (2\pi^2k_{B}m^*)/(\hbar eB)$. $2\pi \beta$ is the Berry phase, and $T_D$ is the Dingle temperature. $\delta$ is a phase shift. $\delta=0$ and $\pm1/8$ for 2D and 3D systems, respectively. The thermal factor $R_T=(\lambda T)/sinh(\lambda T)$ in LK formula has been employed to describe the temperature dependence of FFT amplitude. As shown in the inset of Fig. 5(c), the temperature dependence of relative FFT amplitude can be well fitted, and the extracted small effective masses are listed in Table I, presenting the characteristic of Weyl fermion. Since the previous work has verified that PdGa is a topological semimetal with new types of fermions\cite{schroter2020observation}, the Berry phase from quantum oscillations is not presented. In fact, Berry phases are extracted by the LL index fan diagram while Landau indices n are so large that inevitable error in the fitting is introduced. Thus, we made no further analysis about Berry phase to re-verify the topological properties of PdGa in this work.

The calculated band structure and FSs of PdGa with SOC in Brillouin zone are dispalyed in Figs. 6(a) and (b). According to the first-principles calculations, the $\Gamma$ and R points hold electron-like FSs while the rest hold hole-like FSs. In order to get a complete picture of the fermi surface, detailed cross sections with different value of k$_z$ in the direction of [001] were calculated. Figs. 6(c)-(e) exhibit ten extreme cross sections with B//[001] in different k$_z$ plane.  S$_1$ and S$_2$ are extreme cross sections of two small drop-shaped pockets along $\Gamma$-R cutting at  k$_z$=0.16$\pi$. Cross sections S$_3$-S$_6$ are projections of the FSs around $\Gamma$ point and S$_7$-S$_{10}$ show a quarter of FS$^{'}$s extreme cross sections at M point on k$_z=0$ plane. The calculated FSs in (011) plane are exhibted in Fig. 6(f). According to the Onsager relation, when B//[001], the corresponding oscillation frequencies of S$_1$-S$_6$ are 184.4 T, 227.4 T, 291.6 T, 334.8 T, 382.4 T, 415.1 T, which corresponds to the frequencies F$_1$-F$_6$ as shown in Table I, respectively. At k$_z$=0.2 $\pi$, there is a square extreme cross section in the center of the plane with calculated frequency about 478.5 T, corresponding to the frequency F$_7$ (483.8 T). The calculated frequency of S$_7^{'}$ (4 $\times$ S$_7$) is 553.3 T, which is comparable to the obtained frequency F$_8$ (590.4 T). In a strong magnetic field, carriers can be exchanged between two orbits through a small gap, which is called magnetic breakdown\cite{shoenberg2009magnetic}. A touch of S$_8$ and S$_9$ along M-X is quite apparent, which enables the carriers to move between two orbits in a strong field. Therefore, the corresponding frequency of the difference between S$_8^{'}$ (4$\times$S$_8$), S$_9^{'}$ (4$\times$S$_9$) is 653.1 T in accord with the observed frequency F$_9$ (693.9 T). 

\begin{figure}[tbp]
	\centering
	\includegraphics[width=0.5\textwidth]{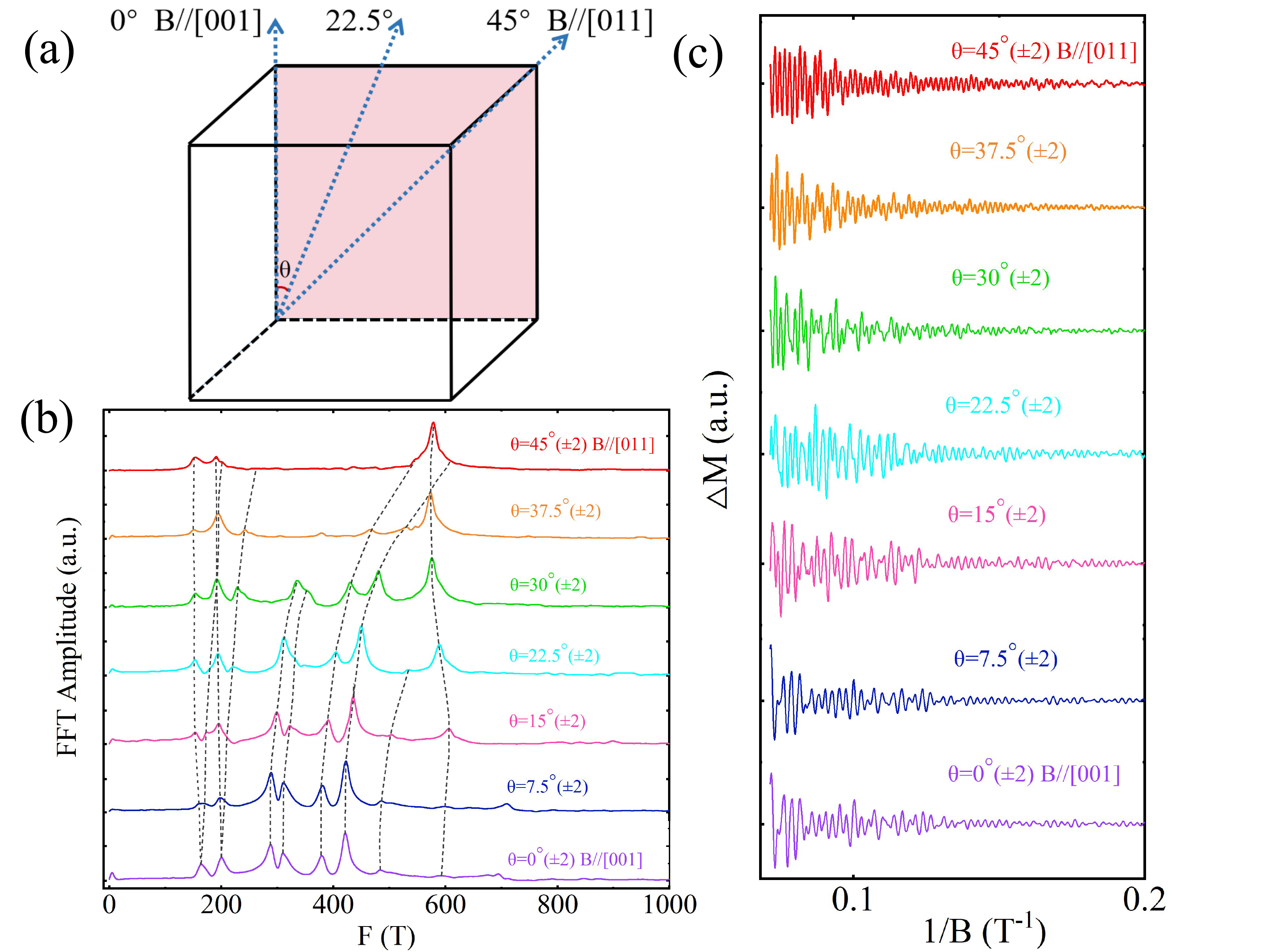}
	\caption{(a) The sketch map of the respective rotating B direction along crystallographic direction. (b) Angle-dependent dHvA oscillations at 2.5 K. (c) Corresponding FFT spectra for B rotating from B//[001] to B//[011].
	}
\end{figure} 

In order to acquire more particular properties of the FSs, the angle-dependent dHvA oscillations have been measured, as displayed in Fig. 7. The schematic diagram in Fig.  7(a) presents the rotation of the magnetic field where $\theta$ is the angle between the magnetic field and [011] direction. We selected seven distinct angles between [001] and [011] for measurement with the field always within (001) plane. As shown in Figs. 6(c) and (f), there are two drop-shaped extreme cross sections near $\Gamma$ point with B//[001] and B//[011]. It shows a good agreement with the results of angle-dependent dHvA oscillations that F$_1$ and F$_2$ gradually split into four peaks as the magnetic field B rotates from [001] to [011], which implies that two FSs evolve into four with different sizes, respectively. And frequencies F$_3$ and F$_4$ fade away at 37.5$^\circ$, which exhibits the feature of the bone-shaped FSs as shown in Figs. 6(b). Two square shaped extreme cross sections at $\Gamma$ point with B//[001] becomes rectangular $\sqrt{2}$ times larger when $\theta$= 45$^\circ$ as shown in Figs. 6(d) and (f), which corresponds to the experimental results that F$_5$ and F$_6$ increase as $\theta$ increases. Similar with the evolutionary trend of the F$_2$ and F$_3$, F$_7$ gradually increases and disappears when $\theta$ = 30$^\circ$, which is in accord with the feature of the endocyclic FS. A larger peak F$_8$ decreases slightly when B turn to [011] direction, which coincides with the extreme cross sections at M point as shown in Fig. 6(d) and (f). As discussed above, magnetic breakdown occurs between two touched FSs of S$_8$ and S$_9$ with B//[001], so it is easily to be destroyed as magnetic field deflect a little bit from [001], corresponding to the change of F$_9$ which is soon disappeared as $\theta$ exceeds 7.5$^\circ$. Thus, we confirm that the frequencies F$_1$, F$_2$, F$_3$, F$_4$ and F$_7$ stem from the FSs along $\Gamma$-R and $\Gamma$-M, while the frequencies F$_5$ and F$_6$, F$_8$ and F$_9$ originate from the FSs at $\Gamma$ and M point.

\section{Summary}

In summary, the high quality single crystals of PdGa with Bi flux are synthesized, which shows a metallic behavior. The large unsaturated MR and Hall resistivity indicate that PdGa should be considered as a multi-band system. Angle-dependent resistivity study reveals weak anisotropy of twofold symmetry or threefold symmetry with B rotating in (011) or (111) plane, respectively, corresponding to the symmetry of the FSs. Band splitting due to SOC in PdGa has been observed in quantum oscillations, which is consistent with the band structure and FSs obtained from first-principles calculations. The frequencies extracted from dHvA oscillations are well consistent with the FSs' extreme cross sections of first-principles calculations.

\section{Acknowledgments}

This work is supported by the National Key R\&D Program of China (Grant No. 2019YFA0308602), the National Natural Science Foundation of China (Grant Nos. 12074425, 11874422), and the Research Funds of Renmin University of China (Grant No. 19XNLG18). K. Liu is supported by the National Key R\&D Program of China (Grant No. 2017YFA0302903), and the National Natural Science Foundation of China (Grant No. 11774424). Computational resources were provided by the Physical Laboratory of High Performance Computing at Renmin University of China.
\bibliography{Bibtex}
\end{document}